# Driver Digital Twin for Online Prediction of Personalized Lane Change Behavior

Xishun Liao, *Student Member*, *IEEE*, Xuanpeng Zhao, Ziran Wang, *Member*, *IEEE*, Zhouqiao Zhao, *Student Member*, *IEEE*, Kyungtae Han, *Senior Member, IEEE*, Rohit Gupta, Matthew J. Barth, *Fellow*, *IEEE*, and Guoyuan Wu, *Senior Member*, *IEEE*

*Abstract*— Connected and automated vehicles (CAVs) are supposed to share the road with human-driven vehicles (HDVs) in a foreseeable future. Therefore, considering the mixed traffic environment is more pragmatic, as the well-planned operation of CAVs may be interrupted by HDVs. In the circumstance that human behaviors have significant impacts, CAVs need to understand HDV behaviors to make safe actions. In this study, we develop a Driver Digital Twin (DDT) for the online prediction of personalized lane change behavior, allowing CAVs to predict surrounding vehicles' behaviors with the help of the digital twin technology. DDT is deployed on a vehicle-edge-cloud architecture, where the cloud server models the driver behavior for each HDV based on the historical naturalistic driving data, while the edge server processes the real-time data from each driver with his/her digital twin on the cloud to predict the lane change maneuver. The proposed system is first evaluated on a human-in-the-loop co-simulation platform, and then in a field implementation with three passenger vehicles connected through the 4G/LTE cellular network. The lane change intention can be recognized in 6 seconds on average before the vehicle crosses the lane separation line, and the Mean Euclidean Distance between the predicted trajectory and GPS ground truth is 1.03 meters within a 4-second prediction window. Compared to the general model, using a personalized model can improve prediction accuracy by 27.8%. The demonstration video of the proposed system can be watched at https://youtu.be/5cbsabgIOdM.

## I. Introduction

### A. Motivation

The rapid development of connected and automated vehicles (CAVs) provides a new perspective on addressing the safety, mobility, and environmental sustainability issues of our transportation systems [1]. However, our transportation systems cannot achieve full automation/connectivity in a foreseeable future, where CAVs have to interact with human-driven vehicles in a mixed traffic environment. To deal with the complex interactions, CAVs need to predict human-driven vehicles' behaviors, make decisions in response to the actions (in presence or to be taken), and execute the right maneuvers through the planner and controller. Particularly, the prediction of human-driven vehicles' behaviors is challenging due to the uncertainties of human drivers. As a result, many researchers [2-5] include driving behavior modeling in the prediction and planning stage, where driver type classification and identification play an important role. Nevertheless, the improvement of integrating the collective driving behavior is limited, because driving behavior can be diverse among different drivers. Therefore, there is an increasing amount of literature that recognizes the importance of personalized driving behaviors.

To study personalized driving behavior, this paper deploys the Driver Digital Twin (DDT) in the real world. DDT is a digital replica of a driver with his or her naturalistic driving data and driving behavior models. Based on real-world data, DDT system in the virtual world provides both online and offline micro-services, e.g., interactive prediction, driving style analysis, etc. In addition, to capture the driving preference variation, the evolving driver model will be updated in a certain period (e.g., every five new trips) by consuming the driving data from the real-world vehicle.

In our previous work, the vehicle digital twin is built on edge server [6], allowing the planning and control for the connected vehicles. However, only using edge server is sometimes insufficient to fulfill the requirements of personalized behavior study, such as data storage, modeling, learning, simulation, and prediction [7]. Therefore, DDT in this study is performed on a cloud-edge architecture, leveraging cloud computing and personalized profiling, enabling both real-time and bulk-batch ingestion, processing and analytics of personal data.

As a typical example of personalized driver behavior that can be modeled by DDT, lane change is a fundamental maneuver of our daily driving and is also among the trickiest maneuvers. Online prediction of the lane change maneuver provides inputs to the downstream motion planners and controllers, allowing CAVs to better cooperate with surrounding human-driven vehicles. By building the personalized lane change model, CAVs can have a better understanding of the specific human drivers and provide a more accurate prediction.

This study was funded by Toyota Motor North America, InfoTech Labs, (Corresponding author: Xishun Liao.)

Xishun Liao, Xuanpeng Zhao, Zhouqiao Zhao, Matthew J. Barth and Guoyuan Wu are with CE-CERT, University of California at Riverside, Riverside, CA, 92507 USA (e-mail: xliao016, xzhao094, zzhao084@ucr.edu, barth@ece.ucr.edu, and gywu@cert.ucr.edu).

Ziran Wang is with the College of Engineering, Purdue University at West Lafayette, IN, 47907, USA (ziran@purdue.edu).

Kyungtae Han and Rohit Gupta are with Toyota Motor North America, InfoTech Labs at Mountain View, CA, 94043, USA (kyungtae.han, rohit.gupta @toyota.com)

In this study, learning-based algorithms for the personalized behavior modeling and online lane change prediction are developed, and field implementations are carried out on a customized vehicle-edge-cloud platform under the digital twin framework. In the field implementation, the cloud (i.e., Amazon Web Services) analyzes the personalized behavior of human-driven vehicles with connectivity (e.g., by cellphone) and stores the learned driver models and historical data, while the edge (i.e., local server) is responsible for the computation of online lane change prediction.

Compared to the existing literature on prediction and behavior modeling, this study has the following main contributions:
- A hierarchical learning-based system is developed for personalized driving behavior modeling, online lane change prediction, and trajectory likelihood estimation.
- Under the digital twin framework, a vehicle-edge-cloud platform is constructed and demonstrated, enabling real-world data collection and algorithms development.
- Personalized driver models are trained and validated using the vehicle-edge-cloud platform, as one of the first real-world deployment of Driver Digital Twin in transportation.
- To validate the proposed algorithm in the field experiments, a portable vision-based human-machine interface (HMI) system is designed to provide the prediction information to the driver supported by edge computing and cloud computing.

### B. Specifications and Assumptions

In this paper, the target predicted vehicle is a connected human-driven vehicle, whose historical/real-time data and the trained driver model are accessible through the digital twin. When other connected vehicles detect and recognize this target vehicle, they can download the driver model of the target vehicle to assist the prediction. Specifically, our prediction algorithm is designed for the on/off-ramp scenario to predict the maneuver and trajectory of the on-ramp vehicle.

To expedite field implementation, some reasonable specifications and assumptions are made in the current stage of online lane change prediction and personalized driving behavior modeling, as follows:
- When the target vehicle comes into view, if ego vehicle is able to recognize it (e.g., by computer vision or V2X communications), the associated driver model is then acquired from the cloud server. If the target vehicle cannot be recognized in time, a general driver model will be used as a backup plan.
- The driving preference of the same driver is assumed to be long-lasting and will not be affected by the mood on the testing day.
- As designing the perception system (e.g., LiDAR, radar, and camera) is outside the scope of this paper, the necessary vehicle information (e.g., location and yaw angle) is uploaded by the target vehicle and shared by edge server.

### C. Organization of the Paper

The remainder of this paper is organized as follows: In Section II, state-of-the-art research on lane change prediction, personalized driver behavior modeling and digital twin platform design are reviewed. Section III explains the methodology of the proposed system, including personalized behavior modeling, online lane change prediction, and map matching for real world implementation. Section IV elaborates on the field experiment design using the vehicle-edge-cloud architecture. In Section V, we analyze the driving styles of two drivers and evaluate the performance of the proposed algorithm. Finally, the paper is concluded with future directions in Section VI.

## II. BACKGROUND AND RELATED WORK

### A. Lane Change Prediction

Hidden Markov Model (HMM) was widely used to infer the lane-change intention [8][9][10] and is usually integrated with the Bayesian network [11] to recognize the lane-change behavior. As the lane-change intention prediction can be modeled as a classified problem, the multilayer perceptron (MLP) was used as a discriminator [12] in long-term lane-change prediction. Moreover, deep learning methods, such as Long Short-Term Memory (LSTM) model, achieved a precision of 90.5% on time-series problems [14]. To find out relevant features for lane changing in a time series, Scheel et al. [15] integrated a temporal attention mechanism with LSTM to improve the prediction accuracy to 92.6% and provide understandability on feature importance. Besides lane change intention prediction, lane change trajectory prediction is also a critical problem. By adding information on traffic level and vehicle types, Xue et al. [16] adopted XGBoost for lane change decision prediction and LSTM for trajectory prediction, achieving a Mean Square Error of 6.62m for trajectory prediction.

However, most of the algorithms ignored the vehicular interaction with the surroundings. Furthermore, supervised learning methods were limited by the lack of enough labeled datasets to cover each possible scenario, and very little research carried out online validation as well as real-world validation.

### B. Personalized Driver Behavior Modeling

A personalized driver model is usually learned based on the dataset from a specific driver (i.e., learning from demonstration) and is mainly used for prediction, planning, and control [17]. Drivers' preferences in their vehicle states are well studied. A personalized driving assistant system developed by Lefevre et al. could identify the current driving maneuver and predict the steering and acceleration, facilitating control assistance [18]. Considering the occupants' preference for lateral and longitudinal comfort, Bae et al. proposed a personalized control system enabling autonomous vehicles to drive like human beings [19].

Not only the vehicle states but also the surroundings and vehicular interaction should be considered in drivers' preference. To integrate the interaction into behavior modeling,

Huang et al. included the awareness of the effect of the ego vehicle on the surrounding vehicles into the cost function [20]. The driver's preference over vehicle states and interactions can be expressed by the cost function recovered by inverse reinforcement learning (IRL), which assumed that human behavior was motivated by optimizing the unknown reward function [21]. In [22], the interaction behavior under different conditions was formulated as a cost function with different linear combinations of features and learned by continuous IRL. The cost function with interpretable weights on features opens the black box of behavior modeling, by showing the diversity of feature attention in different scenarios.

Collecting a personalized dataset for interaction study is another major limitation in interaction modeling and driving personalization study. The difficulty of scenario reproduction constraints the driver experiences in similar interactive conditions and collection of enough data for model training.

### C. Digital Twins for Connected and Automated Vehicles

The recent emergence of the digital twin technology has attracted a significant amount of attention from both academia and industry. The global digital twin market size was reported to be valued at $5 billion in 2020, and will be expanded to $86 billion in 2028, with a compound annual growth rate of 43%. Among all end-uses like manufacturing, energy, and health care, the automotive and transportation industry took one of the largest shares in the global digital twin market in 2020 [23].

By a widely adopted definition (with some variations), a digital twin is a digital replica of a living or non-living physical entity [7]. This concept got to be known by most people in the early 2010s, when NASA adopted it as a key element in its technology roadmap [24]. During the past few years, digital twins have been applied to different vehicular systems. Particularly, Chen et al. developed a "Driver Behavior Twin" to allow driver behavioral models to be shared among multiple connected vehicles to predict future actions of surrounding vehicles [25]. Although this study did not come up with a solid network architecture, its concept did inspire a series of subsequent studies by the authors of our paper.

In 2020, the authors first proposed a digital twin paradigm for an advanced driver-assistance systems (ADAS) with a cloud architecture, which enables the communication between real vehicles and their digital twins deployed on the cloud server in real time [26]. This cooperative ramp merging ADAS was later validated in a field implementation with real passenger vehicles and a private cloud server at University of California, Riverside [6]. Later, the authors introduced edge computing to this network architecture by proposing a mobility digital twin framework, which include not only vehicle digital twins but also human and infrastructure digital twins [7]. Some of the detailed aspects of digital twins for CAVs have also been studied by the authors, such as how to visualize the digital twin information [27], how to leverage the digital twin information for cooperative driving scenarios [28], and how to build a simulation environment to model digital twins [29]. Many of the aforementioned studies have been summarized in a survey paper, where the role of digital twins in CAVs is also compared with the roles of several similar technologies, such as iteration, model-based design, and parallel driving [30].

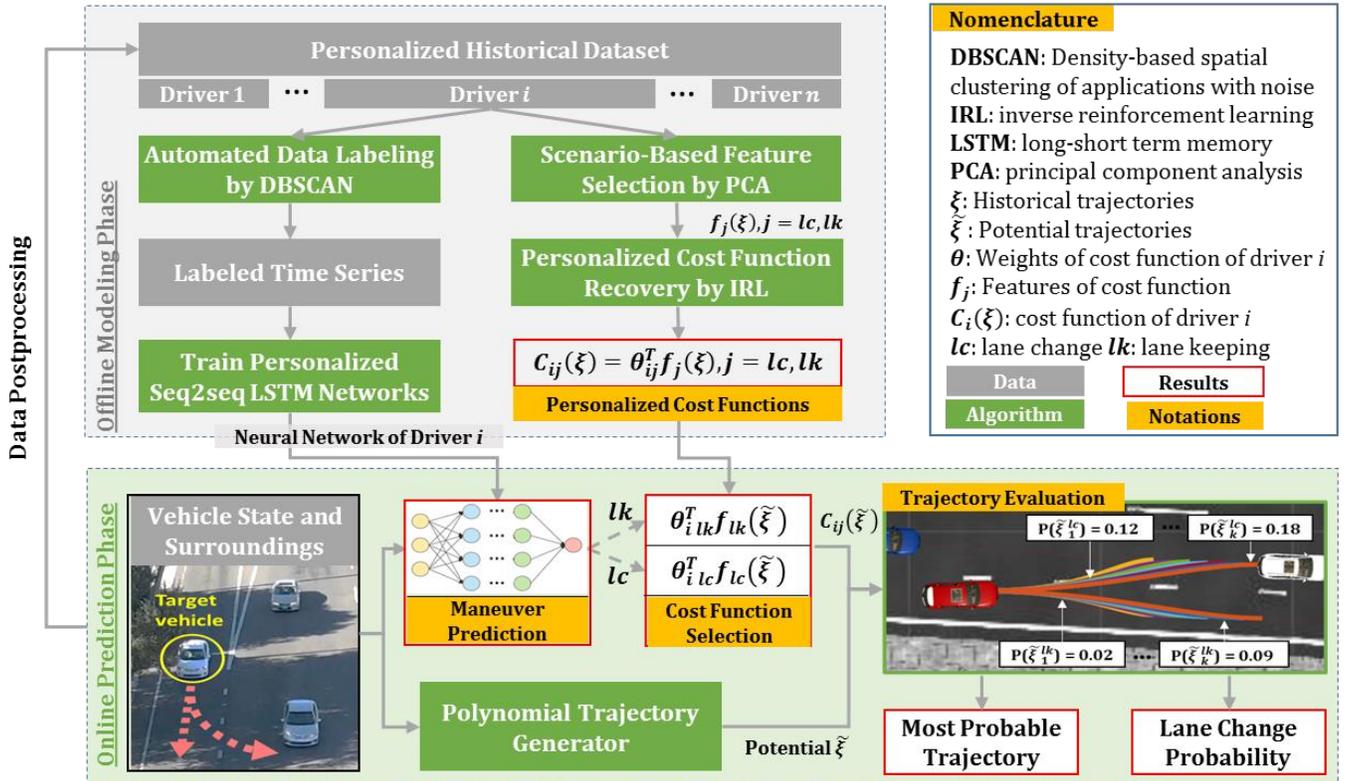

Fig. 1 Personalized lane change behavior modeling: offline learning and online prediction

## III. METHODOLOGY

### A. Personalized Behavior Modeling Process

Based on our previous work [31], we extend the lane change prediction algorithm into a real-world implementable version and enable personalized behavior modeling. As shown in Fig. 1, the algorithm consists of an offline learning process and an online prediction process. The personalized driving behavior for each driver is learned in the offline phase based on the personalized dataset collected from the specific driver. For each driver, a neural network structure sequence-to-sequence (Seq2seq) structure [32] based on LSTM is adopted to predict the lane-change intention, while the cost functions inferring the driver preference are learned by IRL for evaluating the prediction. In the online prediction process, at each time step, the personalized Seq2seq neural network recognizes the maneuver and selects a proper cost function. Then, the driver's cost function evaluates the probability of candidate trajectories provided by the trajectory generator. Finally, the system outputs are the most probable trajectory and the respective lane change probability.

### B. Lane-Change Maneuver Prediction

In order to analyze the trajectory in detail, we need to recognize the driver's intention. We assume that before planning the trajectory, the human driver first considers high-level tasks (e.g., lane change and lane keeping). Therefore, the lane-change intention prediction is formulated as a time-series classification problem, which predicts vehicle states in future time steps, i.e., either lane change or lane keeping. That is to classify the future $T$-step actions $A_{t:t+T}$ into $\{a_{\text{change}}, a_{\text{keep}}\}$, given historical vehicle states and the map information.

To model long-term temporal dependencies among time series, LSTM network is chosen, as its time series prediction capability the is validated in many existing studies, e.g., [33][34][35]. Since each vehicle state in the time series is highly correlated with its adjacent time steps, the Seq2seq neural network using two LSTMs [36] is adopted for a multi-step and multivariable prediction. The neural network input is a trajectory sequence $\xi = (s_{t-T+1}^{lstm}, \ldots, s_t^{lstm})$ of the last $T$ steps, and the vehicle states $s_t$ consists of yaw angle, lateral speed, longitudinal speed, and remaining distance for a mandatory lane change. The output is the predicted lane-change action sequence $(A_{t+1}, \ldots, A_{t+T+1})$ for the next $T$ steps.

### C. Probability Estimation and Trajectory Prediction Based on IRL

The driver behavior and preference are usually represented by the cost function, and rational drivers are assumed to behave for optimizing their cost functions. Considering the continuity of the trajectory space, this study adopts Continuous IRL with Locally Optimal Examples [37][38] to recover this unknown cost function from expert demonstrations.

#### 1) Continuous IRL

The cost function is a linear combination of a set of features, i.e., $C_i(\theta_i, \xi) = \theta_i^T f_i(\xi), i = a_{\text{change}}, a_{\text{keep}}$, where $\theta_i^T$ is the weights vector emphasizing the features, and $f_i(\xi) = \|f_i(s_1, s_2, \ldots, s_t)\|_2$. The goal of the IRL is to figure out the optimal weights $\theta_i^*$ to describe each driver's preference, which maximizes the likelihood of the driver's historical trajectories $\Xi = \{\xi_k\}$, shown in (1):

$$\theta_i^* = \arg \max_{\theta_i} P(\Xi \mid \theta_i) \tag{1}$$

According to the principle of maximum entropy, as shown in (2), a trajectory with a low cost has a higher probability, which is proportional to the exponential of its cost.

$$P(\xi \mid \theta_i) = \frac{e^{-C_i(\theta_i, \xi)}}{Z(\theta)} = \frac{e^{-\theta_i^T f_i(\xi)}}{\int e^{-\theta_i^T f_i(\tilde{\xi})} d\tilde{\xi}} \tag{2}$$

where $Z(\theta) = \int e^{-\theta_i^T f_i(\tilde{\xi})} d\tilde{\xi}$ is the partition function integrating all arbitrary trajectories $\tilde{\xi}$. To handle the computational complexity in solving the partition function, the continuous IRL approximates (1) and reformulates the problem as a minimization of $-\log P(\Xi \mid \theta_i)$:

$$\theta_i^* = \arg \min_{\theta_i} \sum_{k=1}^{K} \frac{1}{2} \mathbf{g}_{\theta_i}^T(\xi_k) \mathbf{H}_{\theta_i}^{-1}(\xi_k) \mathbf{g}_{\theta_i}(\xi_k) - \frac{1}{2} \log |\mathbf{H}_{\theta_i}(\xi_k)| \tag{3}$$

where $\mathbf{g}^T$ and $\mathbf{H}$ are the gradient and Hessian, respectively. This formula indicates that along the expert demonstration the recovery cost function should have small gradients and large positive Hessians.

#### 2) Cost function feature selection

The selected features present the vehicle state in an interpretable way and can capture the preference of the driver. We select the following features to calculate the cost function, based on the available Inertial Measurement Units (IMU) and Global Navigation Satellite Systems (GNSS) information.

a) Lane-change risk $f_{thw}$: ego vehicle is projected to its adjacent lane and calculates the time headway to its potential leading vehicle and the time headway from its following vehicle.

b) Lane-change urgency $f_{urge}$: If the ego vehicle needs to perform a mandatory lane change, the remaining time distance should be considered.

c) Mobility $f_m$: Drivers have different preferences on mobility, and the difference between current speed and the speed limit ($v_{lim}$) is used to evaluate this preference.

d) Comfort $f_a$ and $f_\omega$: The absolute value of the longitudinal acceleration $a_{lon}$ and the yaw rate $\omega$ is used to gauge comfort preference.

e) Lane deviation $f_{dev}$: We also include lateral distance into the cost function to evaluate the imperfection of driving along the centerline of the lane even in the lane-keeping stage.

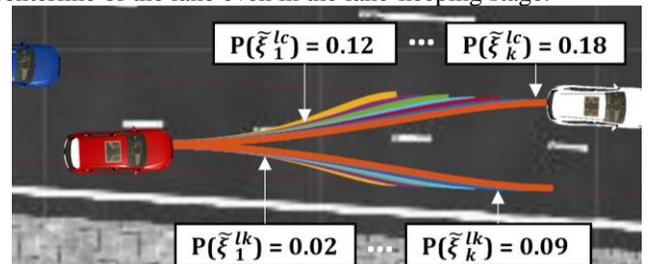

Fig. 2 Polynomial trajectory generator

*3) Trajectory evaluation*

To execute the decision of lane change or lane keeping, planning the trajectory is essential. Considering the real-time performance, instead of exploring arbitrary trajectory, we adopt a polynomial trajectory generator [20] to plan the candidate trajectories $\tilde{\xi}_k$, As in Fig. 2, at each time step, this trajectory generator takes the vehicle's state $\{x, y, v, a, yaw\}$, as inputs and generates multiple trajectories within a prediction window. In this study, we set the planning time window as 4 seconds.

Based on (4), the cost function $C_i(\theta_i, \tilde{\xi}_k)$ is used to evaluate the probability of each possible trajectory $\tilde{\xi}_k$, and select the most probable trajectory. The probability of the lane change maneuver prediction is evaluated by (5), i.e., the probability of lane change equals the sum of the probabilities of all sampled lane-change trajectories.

$$P(\xi_k \mid \theta_i^*) = \frac{e^{-c_i(\theta_i^*, \xi_k)}}{\sum_{k=1}^{K} e^{-c_i(\theta_i^*, \xi_k)}} \quad (4)$$

$$P(\hat{a}_i) = \sum_{k=1}^{K} P(\tilde{\xi}_k \mid \theta_i^*) \quad (5)$$

---

*Algorithm 1*: HMM Map Matching Algorithm

**Input**: 1. Current GPS point ($z_t$). 2. Road segments on map ($r_i$)
**Output**: 1. Matching point ($r_{best}$). 2. Distance to road
1: if (t = 0) Initialize $z_{t-1}$
2: Find candidate road segments ($r_i$) based on $z_t$
3: Calculate the projection $x_{t,i}$ of $z_t$ on each $r_i$
4: Calculate the *measurements probability* $(z_t|r_i)$ by

$$(z_t|r_i) = \frac{1}{\sqrt{2\pi}\sigma_z} exp\left(-0.5\left(\frac{\|z_t - x_{t,i}\|_{great\ circle}}{\sigma_z}\right)^2\right)$$

5: Calculate the *transition probability* $p(d_t)$ by
$$p(d_t) = \frac{1}{\beta} exp\left(-\frac{|\|z_t - z_{t-1}\|_{great\ circle} - \|x_{t,i} - x_{t-1,j}\|_{route}|}{\beta}\right)$$

6: Find the best-fitted road segment by
$$r_{best} = \underset{i}{argmax}(p(z_t|r_i) * p(d_t))$$
7: Find the projection $x_{t,best}$ of $z_t$ on $r_{best}$
8: Calculate the distance between $z_t$ and $x_{t,best}$ by
$$\|z_t - x_{t,best}\|_{great\ circle}$$
9: Update state for next coming measurement $z_{t-1} = z_t$
**Return** $r_{best}$ and $\|z_t - x_{t,best}\|_{great\ circle}$

---

### D. Hidden Markov Model Map Matching

Map matching is an important component in a field implementation to reduce the effect of noisy GPS measurements. Hidden Markov Model (HMM) exploits the road connectivity information and time-sequence feasibility to solve the problem. As shown in *Algorithm 1*, we briefly introduce the HMM map matching algorithm, which was proposed by Newson and Krumm [39]. For the HMM map matching process, we firstly predefine a road network with a set of geographical coordinate pairs (i.e., latitude and longitude) of the road segments based on the real-world situation. The input is the GPS measurement points from the current time step and the last time step. The goal of map matching is to find the best-fitted road segment from all candidate road segments. The candidate road segments are selected by the great-circle distance between the measurement points and projection points on the road segments. Among candidates, we find the one which maximizes the product of the measurement probability and transition probability. Then we output the projection point of the current GPS measurements on the best-fitted road segment as the matching point and the distance between them.

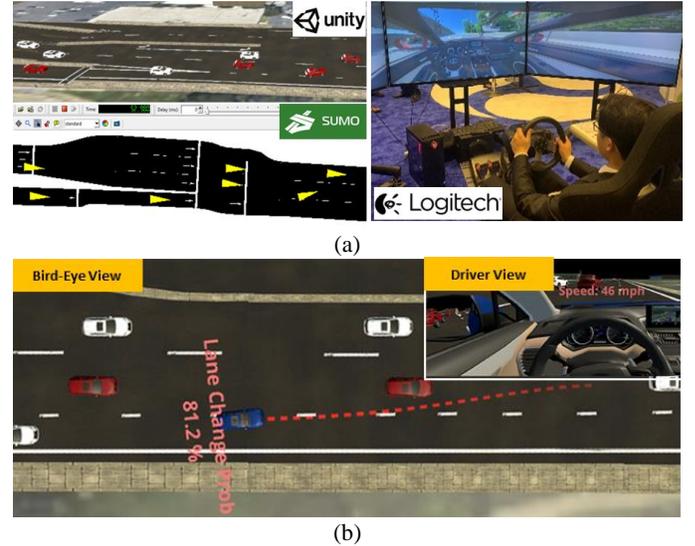

Fig. 3 Algorithm validation on UCR's human-in-the-loop co-simulation platform: (a) major elements in platform. (b) predictions for lane change and most probable trajectory.

### E. Algorithm Validation in Simulation

Before modeling the personalized driving behavior in the real-world, the prediction capability of the proposed algorithm is validated on a general driving behavior model in simulation [31]. In the human-in-the-loop co-simulation platform [40], a real-world test track is programmed in the Unity game engine as the digital version of the real-world testbed to be introduced in Section IV. In the simulation, a mixed traffic flow is generated by SUMO, and human input is consumed by Unity with Logitech driving set, as shown in Fig. 3 (a), allowing various drivers to conduct human-in-the-loop simulations in an immersive traffic environment, where drivers can experience mixed traffic with different CAV penetration rate and congestion levels. To model the general lane changing and lane keeping behavior, 59 trips are collected from ramp drivers within the on-ramp/off-ramp area, under a volume-to-capacity (V/C) ratio of 0.6.

The real-time predicted lane change probabilities and trajectories are visualized in Fig. 3 (b), and the proposed algorithm recognizes the lane change maneuver in 3 seconds before the vehicle crosses the lane separation line. Moreover, the Mean Euclidean Distance (MED) [41] is used to quantify the accuracy of trajectory prediction. This general model achieves a mean MED 0.39 m on average within a 4-second prediction window for 10 test trips, outperforming the IRL-based prediction method in [38], which achieves a mean MED of 0.62 m in a 3-second prediction window. In the field implementation, we further study how the personalized model improves the prediction.

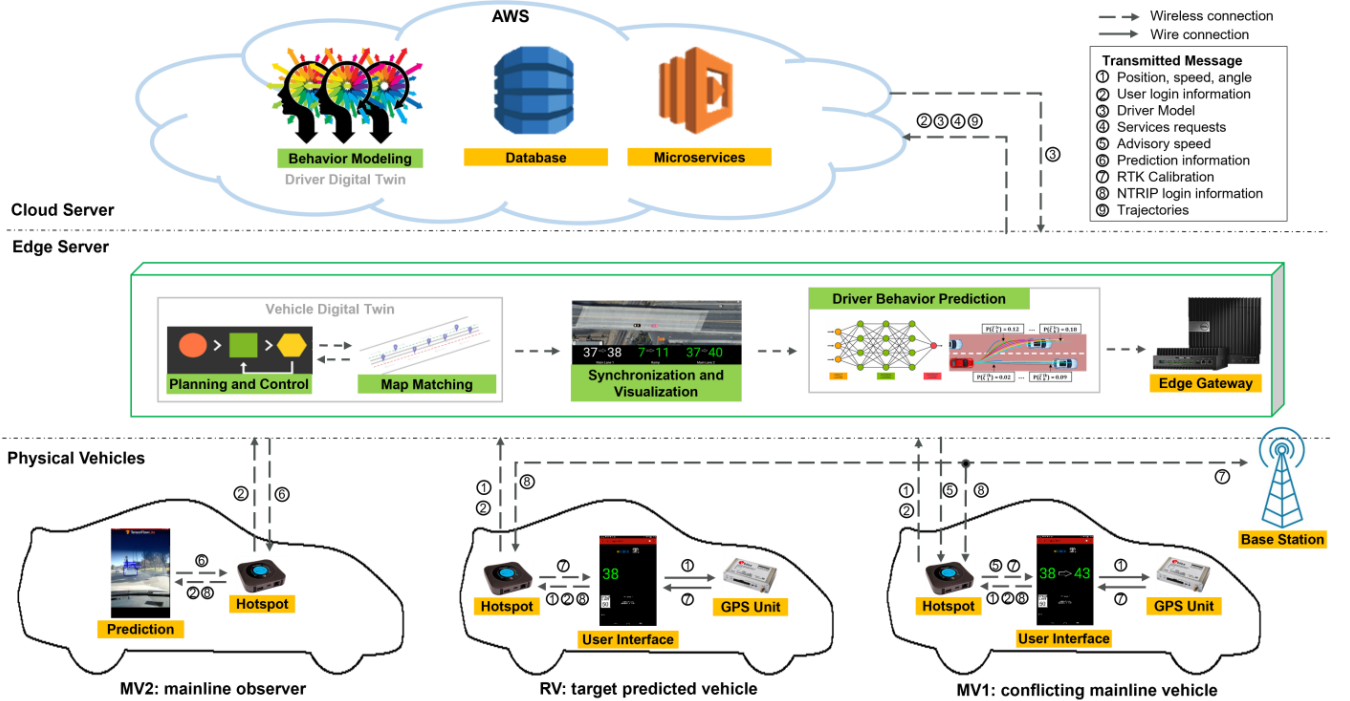

Fig. 4 The vehicle-edge-cloud architecture validated in the field experiment

## IV. DESCRIPTION OF FIELD OPERATIONAL SYSTEM

### A. System Setup

The vehicle-edge-cloud architecture is explained in Fig. 4 including key components, hardware for the real-world implementation, and the information flow between them. Under this architecture, vehicles are considered service consumers who store their personalized dataset on the cloud server and share the real-time perception information with the edge server, from which vehicles receive driving assistance and support services to facilitate automation. The edge server creates the vehicle digital twin and serves as the bridge between the cloud server and vehicles. DDT is created on a cloud server by the offline modeling algorithms described in Fig. 1.

To be specific, the first step of building the vehicle digital twin on the edge server is to know the position of each vehicle in the network. After building the connection between vehicles, the perception information and GPS coordinates of each vehicle are uploaded to the edge server, where the map matching algorithm updates the vehicle positions in real time. Furthermore, algorithms for predictions, planning, and control can be provided by request for different purposes. With the connection to the cloud server, the edge server receives service requests from vehicles, and passes on the request and the unique login information to the cloud. In response to the request, the cloud can feed back edge server with driving behavior models of specific drivers. Due to the high communication latency between the cloud and vehicles, it is hard to implement real-time services provided directly from the cloud to customers. To address this problem, Edge Gateway is adopted to handle the data exchange and provide services to vehicles in real time.

The concern of privacy is also addressed, as only unidentifiable information is transmitted. Cloud server is responsible for driver model training, storage of personalized data and models, and microservices support (e.g., energy consumption analysis). Therefore, we take advantage of the strong computational power, high-speed data processing, and secure data storing features of AWS. Historical personal driving data are archived in the personal folder and can only be accessed by the personal login information. Service consumers can also request post-processed driving behavior reports, (e.g., energy consumption analysis, to better understand or improve their driving skills.

Hardware for the real-world implementation is shown in the Fig. 4. All three vehicles are 2012 Corolla LE models, with 1.8-liter internal combustion engines. Each vehicle is equipped with a Wi-Fi hotspot (Netgear MiFi) to establish a wireless connection with edge server, a GPS unit to collect accurate position information of vehicles, and a portable human-machine interface (HMI) device using Galaxy A7 tablet. The U-blox C102-F9R is adopted as the GPS unit, which is a multi-band GNSS with Real-Time Kinematic positioning (RTK) function and sensor fusion technologies achieving decimeter-level accuracy. The GPS unit and the HMI device are connected (wired) with serial communications. The Wi-Fi hotspot handles the wireless communications between the tablet and other infrastructures such as the base station and edge server. Depending on the request of the driver, the HMI device receives and visualizes the speed guidance or lane-change prediction information from the edge server. Moreover, to enable the RTK function in the U-blox and get high-accuracy GPS measurements, the HMI device shares the calibration correction message received from the base station to its paired GPS unit.

The edge gateway is running on a customized edge server installed at CE-CERT, University of California, Riverside, which is 1.2 km away from the testing site. A Dell R630 server, with two Xeon 2.4GHz (6-core) CPUs, 64GB RAM, 1TB solid state drive and 14 TB hard disk drive, is adopted as the edge server in this implementation.

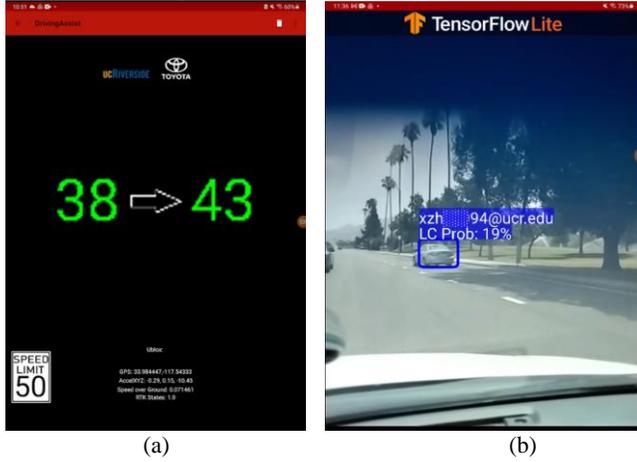

Fig. 5 Vehicle HMI design. (a)Speed guidance, and (b) detection for lane changing prediction

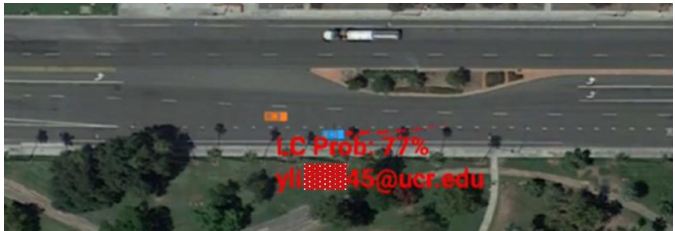

Fig. 6 The bird-eye's view application interface: digital twin of vehicles running on the edge server in real time.

### B. Experiment Plan

As shown in the vehicle level of Fig. 4, the real-world implementation is conducted by three passenger vehicles which are mainline vehicle 1 (MV1), mainline vehicle 2 (MV2), and ramp vehicle (RV), respectively. Specifically, RV is the target predicted ramp vehicle, MV1 is the conflicting vehicle with RV. While MV2 drives behind MV1 and is potentially affected by the lane change maneuver of RV, MV2 observes and predicts the whole lane change process. To ensure MV1 and RV can encounter with each other, speed guidance is provided to MV1 only before reaching an observable point, from which MV1 and RV can see each other for the rest of the trip. The speed guidance [6] is calculated by edge server, where RV has its virtual projection on the mainline, and MV1 is assigned to follow the virtual projection without an intervehicle gap. By following the speed guidance, MV1 can reach the observable point at the same time as RV, and for each experiment, nearly the same condition of merging conflict is guaranteed. Different from MV1, to collect naturalistic driving data of the RV driver, RV receives no interference from the system during the whole trip. During the conflict creation process, MV2 is not involved since it plays the role as an observer and has no impact on the lane change. If MV2 successfully detects and recognizes RV, MV2 sends request and downloads the driver behavior model of RV from the cloud to facilitate the lane change prediction.

To meet the above demands, the HMI guides MV1 with the speed guidance which shows the current speed on the left and the target speed on the right as shown in Fig. 5 (a). For MV2, we leverage the benefits from TensorFlow Lite Object Detection, an open-source Android application, to detect the RV and visualize the predicted lane changing probability with user ID on the top of a bounding box as indicated in Fig. 5 (b).

Before starting the vehicle, each driver logs in with a unique ID and password to access his/her personalized digital twin. In addition, to obtain personalized services, drivers need to agree with sharing information with edge server.

A bird-eye view Android application is designed for visualizing the vehicle Digital Twins running on the edge server. MV1 and RV are the orange vehicle and the blue vehicle respectively in Fig. 6. Moreover, the lane changing probability and the predicted trajectory of RV are provided for a better understanding of the entire field implementation process.

To build DDT on cloud server, personalized dataset for each RV driver is collected. In total, 20 trips entering the mainline and 20 trips driving off-ramp are used for behavior modeling. The average duration of each trip is 35 seconds, with an average update rate of 5Hz. In each time step, both the trajectories of MV1 and RV are recorded and synchronized at edge server, and the dataset on the cloud server is updated at the end of the trip.

TABLE I
WEIGHTS OF RECOVERED COST FUNCTIONS

| Scene | Driver | $f_a$ | $f_{thw}$ | $f_{dev}$ | $f_{yaw}$ | $f_m$ | $f_{urge}$ |
|---|---|---|---|---|---|---|---|
|    | #1 | 0.696 | 0.151 | 0 | 0 | 0.023 | 0.126 |
| LC | #2 | 0.425 | 0.240 | 0 | 0 | 0.182 | 0.151 |
|    | General | 0.529 | 0.284 | 0 | 0 | 0.179 | 0.008 |
|    | #1 | 0.388 | 0 | 0.246 | 0.348 | 0.017 | 0 |
| LK | #2 | 0.323 | 0 | 0.222 | 0.043 | 0.412 | 0 |
|    | General | 0.356 | 0 | 0.311 | 0.309 | 0.023 | 0 |

## V. RESULT ANALYSIS

### A. Cost function and driving pattern analysis

Personalized models are trained using drivers' own datasets (i.e., 19 trips collected from Driver 1 and 20 trips from Driver 2). For two drivers, while a general model is trained using the aggregated dataset from both drivers, standing for a collective driving behavior. The weights in the cost functions can reflect drivers' preference when large weights penalize high values, as shown in Table I. For lane change maneuvers, Driver 1 cares more about longitudinal comfort and penalizes large acceleration with 0.696 on $f_a$, compared with 0.425 of Driver 2. Driver 2 prefers a smaller time gap during lane change as he/she puts a larger weight on $f_{thw}$ than driver 1 to penalize the time headway. Driver 1 does not care about the mobility $f_m$ and drives slower than the speed limit, while Driver 2 prefers to drive faster with a weight of 0.182. For lane keeping maneuvers, two drivers show the same preference for longitudinal comfort $f_a$ but significant differences in lateral comfort and mobility. Driver 1 tends to keep the vehicle stable (i.e., putting a large weight on $f_{yaw}$ penalizing unstable yaw movements) and drives slowly. On the contrary, Driver 2 pays little attention to the stability of the vehicle pose and seeks high speed. According to the analysis of cost function weights, Driver 1 is more likely to be a cautious driver while Driver 2 tends to be more aggressive. For lane changing, general model puts least weight on urgency $f_{urge}$ and penalizes large time headway. For lane keeping, general model is sensitive to lane deviation $f_{dev}$ the most.

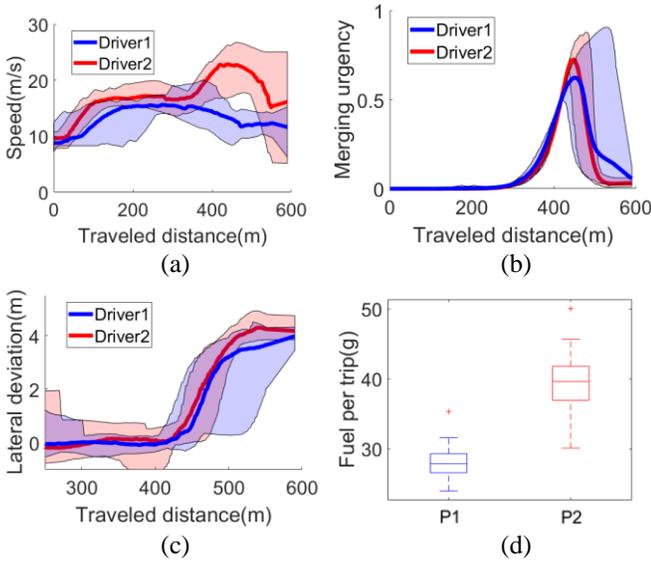

Fig. 7 Driving pattern comparison during a lane change process: (a) longitudinal speed of ramp vehicle (b) mandatory lane change urgency (c) lateral behavior and (d) fuel consumption.

Besides the cost function analysis, the driving pattern can also be recognized via the overall lane change behavior, as shown in Fig. 7.

The interaction with the mainline vehicle can be reflected by the longitudinal speed of the lane change process in Fig. 7(a), which displays the median speed at each location. The observable point (at 320m) is 100 meters before the lane change point, where mainline and ramp drivers can see each other for the first time during the lane change. After observing the conflict with MV1, Driver 1 chooses to slow down and yield to MV1 for lane change behind, but Driver 2 accelerates to surpass in order to cut in front of MV1. The lane change urgency in Fig. 7(b) is used to measure how the driver deals with a mandatory lane change. The urgency value grows when RV comes closer to the end of the lane-change area without changing the lane. Once the lane change is completed, the urgency will decrease shortly. Driver 1 has a smaller peak value than Driver 2 for lane change urgency; however, the urgency pattern of Driver 2 is more consistent, as the urgency variation (red area) is smaller than the one of Driver 1 (blue area).

The lateral movement preference is captured by the lateral deviation shown in Fig. 7 (c), where the lane change is completed once the lateral deviation reaches 4m or above. Although two drivers have different preferences for lane change sequences, their lane change starting points are close to each other. Moreover, two types of slopes (i.e., lane change speed) in the blue line are observed during the lane change process of the driver. In the first segment, Driver 1 merely crosses the lane separation line, and in the second segment, he/she approaches the center line slowly after confirming safety.

A similar conclusion to cost function analysis can be made that Driver 1 is more cautious than Driver 2. As a result, as shown in Fig. 7 (d), the average fuel consumption of Driver 1 (28.2g) is 28.9% less than the one of Driver 2 (39.7g), in each lane change.

*B. Case study*

As presented in Fig. 8, the online prediction combines probability estimation and trajectory prediction. In stage 1 see Fig. 8(a), RV enters the interacting zone and is recognized by MV2 (observer). At the same time, MV2 sends a request to edge server, and then edge server receives the driver model from the cloud. Then, the driver ID and lane change probability are visualized on both observer's view and edge. Meanwhile, the lane change probability showing above the blue bounding box is low, and the predicted trajectory (visualized on edge in red dash line) is straight, indicating the lane keeping maneuver. In stage 2 see Fig. 8(b), RV slows down to yield MV1, which is captured by MV2. At the moment, the predicted trajectory points to the lane separation line, and the lane change probability increases with the color of the bounding box changing from blue to red for warning of a potential lane change. In stage 3 see Fig. 8(c), Driver 1 perceives that the gap is large enough, and the lane change begins. MV2 is more certain about the prediction, showing a high probability and a predicted trajectory pointing to the center of the mainline. Fig. 8 (d)(e)(f) present a similar process for Driver 2's lane change. One noticeable difference from Driver 1's model is that Driver 2's personalized model pays attention to speeding up rather than slowing down.

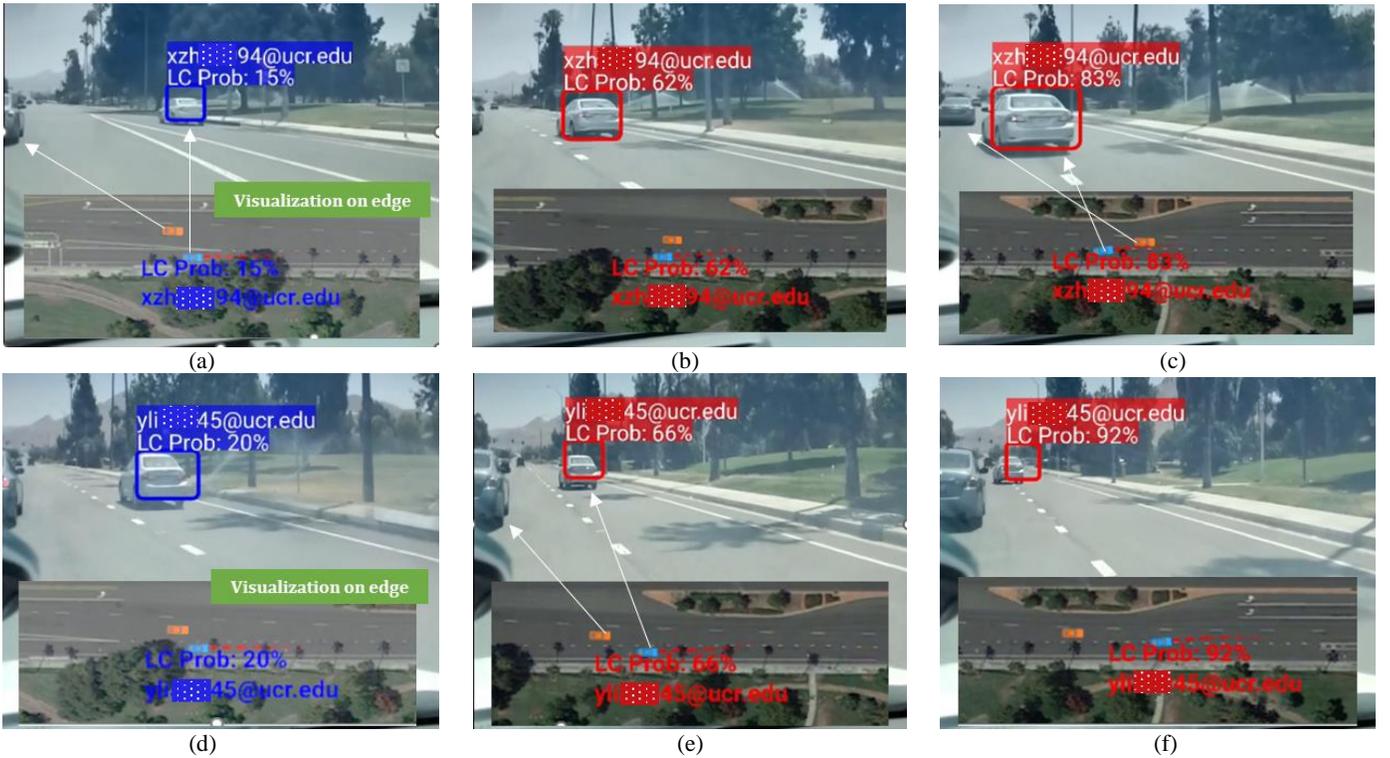

Fig. 8 Field implementation for online lane change prediction with visualization on camera and edge server: (a)(b)(c) Prediction for Driver 1, changing the lane behind MV1. (d)(e)(f) Prediction for Driver 2, changing the lane in front of MV1.

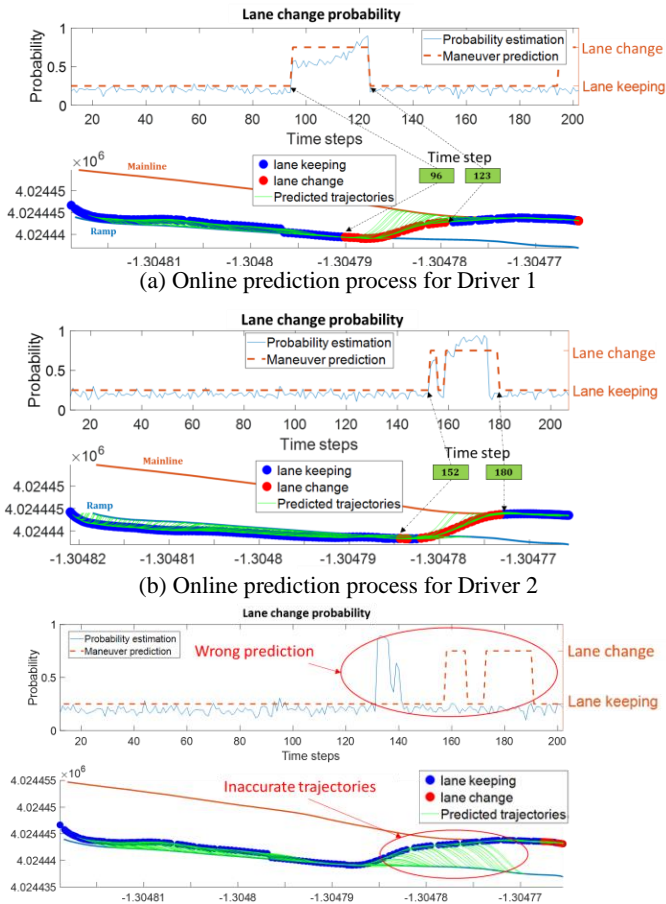

(a) Online prediction process for Driver 1

(b) Online prediction process for Driver 2

(c) Offline prediction using a mismatched driver model

Fig. 9 Prediction result analysis of using personalized models

Fig. 9 elaborates the whole prediction processes of the same trips in Fig. 8. Specifically, Fig. 9 (a) presents the prediction result of one lane change performed by Driver 1. The top subfigure depicts the prediction result and probability estimation of lane change during a trip, reflecting the intention of the driver. In the bottom subfigure, each time step of the ground-truth trajectory is labeled by neural network in real time as lane change (red dots) or lane keeping (blue dots), when the predicted trajectory of each time step is shown in the green line.

The maneuver prediction can be corrected by the probability estimation. In the top subfigure of Fig. 9 (b), when the lane change has been completed, lane change is still predicted by LSTM (orange dash line) for the time steps from 177 to 180, but the probability of lane change (blue solid line) is estimated only 20%. According to the bottom subfigure, Driver 2 has completed the lane change at the 177th step, and the cost function-based probability estimation corrects the neural network prediction.

Personalized models can only be used on a specific person. Fig. 9 (c) illustrates how will the prediction result be when the model is mismatched (using Driver 1's model on Driver 2's trip), where the predictions are not accurate for lane change probability, action, and trajectory.

The prediction needs to be evaluated for both maneuver and trajectory. For maneuver prediction, the predictive capability can be quantified as the time between the moment of recognizing the lane change intention and the moment of the vehicle crossing the lane separation line. The personalized model of Driver 1 recognizes the lane change intention in 6.08 seconds on average, with a 1.96-second standard deviation (STD), while Driver 2's model achieves an average of 3.73 seconds with an STD of 1.29 seconds. The Mean Euclidean Distance (MED) is used to analyze the accuracy of trajectory prediction. Although we cannot decouple the GPS error and prediction error, the model achieves a good result.

The quantitative accuracy evaluation is shown in Table II. Using either general or personalized model, the prediction of Driver 1 is better than Driver 2. The MEDs between predicted trajectory and GPS points are 1.03 m (STD: 0.4 m) for Driver 1 and 1.48 m (STD: 1.05 m) for Driver 2 in a 4-second prediction window. Compared to results from the general model, the personalized model improves Driver 2 's results the most, by 27.8% on average. Since Driver 1 is more predictable, the improvement of using personalized model is limited (by 1.9%), but the prediction variation is reduced by 42%.

TABLE II
ACCURACY COMPARISON FOR GENERAL MODEL AND PERSONALIZED MODEL

|  | MED (m) | General | Personalized | Improvement |
|---|---|---|---|---|
| Driver 1 | Mean | 1.05 | 1.03 | 1.9% |
|  | STD | 0.69 | 0.4 | 42% |
| Driver 2 | Mean | 2.05 | 1.48 | 27.8% |
|  | STD | 1.17 | 1.05 | 10% |

## VI. CONCLUSIONS AND FUTURE WORK

This paper has proposed an online lane change prediction algorithm based on personalized driving behavior modeling, which is validated on a vehicle-edge-cloud testbed under the Driver Digital Twin (DDT) concept. Specifically, a sequence-to-sequence LSTM neural network has been used to predict the lane change intention, and personalized driving behaviors have been modeled for different drivers, whose preferences are learned and analyzed by inverse reinforcement learning based on the historical data stored on the cloud server. Supported by the personalized models, an online lane change prediction system has been developed and validated with real-world field implementation. The system is able to recognize the target driver's lane change intention at 6.08 seconds before the vehicle crosses the lane separation line, and the Mean Euclidean Distance between the predicted trajectory and ground truth (based on the measurements from an RTK-enabled GPS unit) is 1.03 m within a 4-second prediction window. Using a personalized model can improve the prediction accuracy by 27.8%.

As one of the first few research projects looking into personalized driving behavior, there are still some limitations of this implementation, and improvements can be made alongside the future development. The prediction algorithm is specifically designed for on/off-ramp mandatory lane changing. The mechanism behind discretionary lane changing maneuvers can be different, which requires adjustment on feature selection of the cost function. Another major constraint on studying personalized behavior is data availability, which can be relieved by transferring the model learned from simulation. Besides the research on personalized behavior modeling, incorporating personalized prediction with planning is also an important future step, as it allows CAVs to drive like human-driven vehicles, and improve their user acceptance and trust.

## ACKNOWLEDGMENT

This research was funded by Toyota Motor North America, InfoTech Labs under the Digital Twin project. We are grateful to Yejia Liao for his contributions to the field implementation.

The contents of this paper only reflect the views of the authors, who are responsible for the facts and the accuracy of the data presented herein. The contents do not necessarily reflect the official views of Toyota Motor North America.

[17] M. Hasenjäger, M. Heckmann and H. Wersing, "A Survey of Personalization for Advanced Driver Assistance Systems," in *IEEE Transactions on Intelligent Vehicles*, vol. 5, no. 2, pp. 335-344, June 2020.

[18] S. Lefevre, A. Carvalho, Y. Gao, H. E. Tseng, and F. Borrelli, "Driver models for personalised driving assistance," *Vehicle System Dynamics*, vol. 53, no. 12, pp. 1705–1720, 2015.

[19] I. Bae, J. Y. Moon and S. Kim, "Self-driving like a human driver instead of a robocar: Personalized comfortable driving experience for autonomous vehicles," 2020, *arXiv*:2001.03908

[20] Z. Huang, J. Wu, and C. Lv, "Driving behavior modeling using naturalistic human driving data with inverse reinforcement learning," *IEEE Transactions on Intelligent Transportation Systems*, 2021.

[21] B. D. Ziebart, A. L. Maas, J. A. Bagnell, A. K. Dey et al., "Maximum entropy inverse reinforcement learning." in *AAAI*, vol. 8. Chicago, IL, USA, 2008, pp. 1433–1438.

[22] M. Naumann, L. Sun, W. Zhan, and M. Tomizuka, "Analyzing the suitability of cost functions for explaining and imitating human driving behavior based on inverse reinforcement learning," in 2020 *IEEE International Conference on Robotics and Automation (ICRA)*. IEEE, 2020, pp. 5481–5487.

[23] Grand View Research, "Digital Twin market size, share & trends analysis report by end-use (automotive & transport, retail & consumer goods, agriculture, manufacturing, energy & utilities), by region, and segment forecasts, 2021 - 2028," Accessed: 2021-12-09. [Online]. Available: https://www.grandviewresearch.com/industry-analysis/digital-twin-market.

[24] E. Glaessgen and D. Stargel, "The digital twin paradigm for future NASA and US Air Force vehicles," in *53rd AIAA/ASME/ASCE/AHS/ASC Structures, Structural Dynamics and Materials Conference 20th AIAA/ASME/AHS Adaptive Structures Conference 14th AIAA*, 2012, p. 1818.

[25] X. Chen, E. Kang, S. Shiraishi, V. M. Preciado, and Z. Jiang, "Digital behavioral twins for safe connected cars," in *Proceedings of the 21th ACM/IEEE International Conference on Model Driven Engineering Languages and Systems*, 2018, pp. 144–153.

[26] Z. Wang, X. Liao, X. Zhao, K. Han, P. Tiwari, M. J. Barth, and G. Wu, "A Digital Twin paradigm: Vehicle-to-Cloud based advanced driver assistance systems," in *2020 IEEE 91st Vehicular Technology Conference*, May 2020, pp. 1–6.

[27] Y. Liu, Z. Wang, K. Han, Z. Shou, P. Tiwari, and J. Hansen, "Vision-cloud data fusion for ADAS: A lane change prediction case study," *IEEE Transactions on Intelligent Vehicles*, vol. 7, no. 2, Jun. 2022.

[28] Z. Wang, K. Han, and P. Tiwari, "Digital Twin-assisted cooperative driving at non-signalized intersections," *IEEE Transactions on Intelligent Vehicles*, vol. 7, no. 2, Jun. 2022.

[29] Z. Wang, K. Han, and P. Tiwari, "Digital twin simulation of connected and automated vehicles with the unity game engine," in *2021 IEEE 1st International Conference on Digital Twins and Parallel Intelligence (DTPI)*. IEEE, 2021, pp. 1–4.

[30] C. Schwarz and Z. Wang, "The role of digital twins in connected and automated vehicles," *IEEE Intelligent Transportation Systems Magazine*, pp. 2–11, Jan. 2022.

[31] X. Liao, Z. Wang, X. Zhao, Z. Zhao, K. Han, P. Tiwari, M. Barth, and G. Wu, "Online prediction of lane change with a hierarchical learning-based approach," in *Proceedings 2022 IEEE International Conference on Robotics and Automation*, 2022.

[32] I. Sutskever, O. Vinyals, and Q. Le, "Sequence to Sequence Learning with Neural Networks," in *Neural Information Processing Systems*, 2014.

[33] S. Hochreiter and J. Schmidhuber, "Long short-term memory," *Neural computation*, vol. 9, no. 8, pp. 1735–1780, 1997.

[34] F. Altché and A. de La Fortelle, "An LSTM network for highway trajectory prediction," in *2017 IEEE 20th International Conference on Intelligent Transportation Systems (ITSC)*. IEEE, 2017, pp. 353–359.

[35] X. Luo, D. Li, Y. Yang, and S. Zhang, "Spatiotemporal traffic flow prediction with KNN and LSTM," *Journal of Advanced Transportation*, vol. 2019, 2019.

[36] W. Kim, Y. Han, K. J. Kim, and K.-W. Song, ''Electricity load forecasting using advanced feature selection and optimal deep learning model for the variable refrigerant flow systems,'' *Energy Reports*, vol. 6, pp. 2604–2618, 2020.

[37] S. Levine and V. Koltun, "Continuous inverse optimal control with locally optimal examples," *arXiv preprint arXiv*:1206.4617, 2012.

[38] L. Sun, W. Zhan, and M. Tomizuka, "Probabilistic prediction of interactive driving behavior via hierarchical inverse reinforcement learning," in *2018 21st International Conference on Intelligent Transportation Systems (ITSC)*, 2018, pp. 2111–2117.

[39] P. Newson and J. Krumm, "Hidden markov map matching through noise and sparseness," in *Proceedings of the 17th ACM SIGSPATIAL international conference on advances in geographic information systems*, 2009, pp. 336–343.

[40] X. Zhao, X. Liao, Z. Wang, G. Wu, M. J. Barth, K. Han, and P. Tiwari, "Co-simulation platform for modeling and evaluating connected and automated vehicles and human behavior in mixed traffic," *SAE MobilityRxiv™ Preprint*, 2021.

[41] J. Quehl, H. Hu, O. S Tas, E. Rehder, and M. Lauer, "How good is my prediction? Finding a similarity measure for trajectory prediction evaluation," in *2017 IEEE 20th International Conference on Intelligent Transportation Systems (ITSC)*, 2017, pp. 1–6.

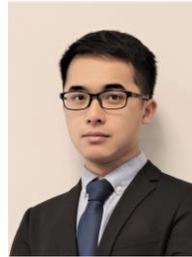

**Xishun Liao** (S'19) received the M.Eng. degree from University of Maryland, College Park in 2018, and the B.E. degree from Beijing University of Posts and Telecommunications in 2016. He is currently a Ph.D. candidate in electrical and computer engineering at the University of California at Riverside. His research focuses on motion planning, driver behavior, and connected and automated vehicle technology.

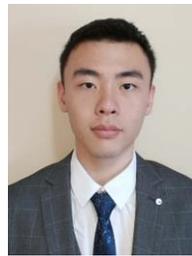

**Xuanpeng Zhao** received the B.E. degree from Shanghai Maritime University in 2019. He is currently a Ph.D. student in electrical and computer engineering at the University of California at Riverside. His research focuses on embedded system, computer vision, and connected and automated vehicle technology.

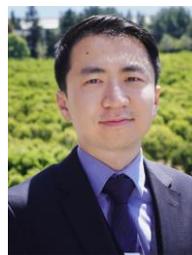

**Ziran Wang** (S'16-M'19) received the Ph.D. degree from The University of California, Riverside in 2019, and the B.E. degree from Beijing University of Posts and Telecommunications in 2015, respectively. He is currently an Assistant Professor at Purdue University, where he leads the Purdue Digital Twin lab. Prior to joining Purdue, Dr. Wang was a Principal Researcher at Toyota Motor North America, InfoTech Labs. Dr. Wang serves as Founding Chair of IEEE Technical Committee on Internet of Things in Intelligent Transportation Systems (IoT in ITS), Associate Editor of IEEE Transactions on Intelligent Vehicles, and Associate Editor of SAE International Journal of Connected and Automated Vehicles. His research focuses on intelligent vehicle technology, including autonomous driving, driver behavior modeling, and vehicular cyber-physical systems.

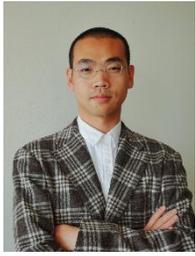

**Zhouqiao Zhao** (S'19) received his M.S degree in Electrical Engineering from Ohio State University in 2017, and the B.S degree from University of Electronic Science and Technology of China in 2015, respectively. He is currently working toward the Ph.D. degree in Electrical Engineering at the University of California, Riverside. He works as a graduate student researcher with the College of Engineering – Center for Environmental Research and Technology, University of California, Riverside. His research focuses on cooperative automation and connected and automated vehicle.

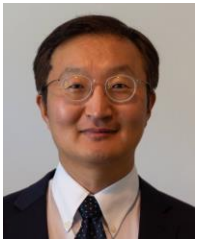

**Kyungtae (KT) Han** (M'97-SM'15) received the Ph.D. degree from the University of Texas at Austin in 2006. He is currently a Senior Principal Scientist at Toyota Motor North America, InfoTech Labs. His research interests include cyber-physical systems, connected and automated vehicles, and intelligent transportation systems.

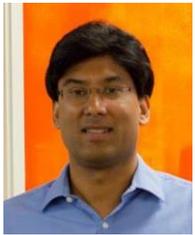

**Rohit Gupta** received the Ph.D. degree in mechanical engineering from The University of California at Santa Barbara, where he did the research for "high speed" ship design and turbulent flow modeling of gas-liquid interfaces. He is currently a Principal Cloud Architect at Toyota Motor North America, InfoTech Labs. Prior to joining Toyota, he worked in various industries: semiconductors, telecommunications, finance, and software. His research interests include cloud and edge computing architectures, connected and autonomous vehicles, AI/ML, "Digital Twin" of human-vehicle interactions, and high-speed mobility.

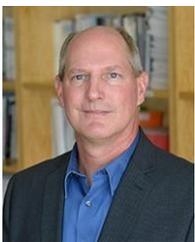

**Matthew J. Barth** (M'90-SM'00-F'14) received the M.S. and Ph.D. degree from the University of California at Santa Barbara, in 1985 and 1990, respectively. He is currently the Yeager Families Professor and Associate Dean of the College of Engineering, University of California at Riverside. His current research interests include ITS and the environment, transportation/emissions modeling, and advanced sensing and control. Dr. Barth has been a Senior Editor for the IEEE Transactions on ITS. He served as the IEEE ITSS President for 2014 and 2015 and is currently the IEEE ITSS Vice President for Educational Activities.

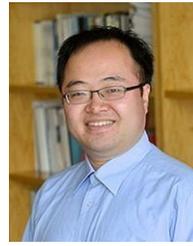

**Guoyuan Wu** (M'09-SM'15) received his Ph.D. degree from the University of California, Berkeley in 2010. Currently, he holds an Associate Researcher and an Associate Adjunct Professor position at Bourns College of Engineering – Center for Environmental Research & Technology and Department of Electrical & Computer Engineering in the University of California at Riverside. His research focuses on development and evaluation of sustainable and intelligent transportation system technologies. Dr. Wu is an Associate Editor of SAE Journal of CAV and IEEE Open Journal of Intelligent Transportation Systems.